\documentclass[9pt,letterpaper]{article}
\usepackage{opex3}
\usepackage{amsmath}
\usepackage{graphicx}
\usepackage{amssymb}
\usepackage{cite}

\begin{document}
\title{Probing rotational wave-packet dynamics with the structural minimum in high-order harmonic spectra}
\author{Meiyan Qin$^{1,2}$, Xiaosong Zhu$^{1,2,}$$^*$, Yang Li$^{1,2}$, Qingbin Zhang$^{1,2}$, Pengfei Lan$^{1,2}$, and Peixiang Lu$^{1,2,3}$}
\address{1 Wuhan National Laboratory for Optoelectronics and School of Physics, Huazhong University of
Science and Technology, Wuhan 430074, P. R. China \\
2 MOE Key Laboratory of Fundamental Quantities Measurement,Wuhan
430074, P. R. China\\} \email{$^*$ zhuxiaosong@mail.hust.edu.cn}
\email{$^3$ lupeixiang@mail.hust.edu.cn}

\begin{abstract}
We investigate the alignment-dependent high-order harmonic
spectrum generated from nonadiabatically aligned molecules around
the first half rotational revival. It is found that the evolution
of the molecular alignment is encoded in the structural minima. To
reveal the relation between the molecular alignment and the
structural minimum in the high-order harmonic spectrum, we perform
an analysis based on the two-center interference model. Our
analysis shows that the structural minimum position depends
linearly on the inverse of the alignment parameter
$<\cos^2\theta>$. This linear relation indicates the possibility
of probing the rotational wave-packet dynamics by measuring the
spectral minima.
\end{abstract}


\section{Introduction}
The main impetus to the development of strong field physics is
measuring and understanding the electronic structure and dynamics
of matter on its natural timescale
\cite{itatani,vozzi,zhu,haessler,qin,torres,zhu0,liao}. When
measuring the electronic structure and dynamics for molecules, the
pre-alignment of the target molecules is required. This can be
adiabatically or nonadiabatically achieved by using moderately
intense laser pulses \cite{torres1,ortigoso,seideman}. Generally,
the nonadiabatic alignment is preferred because it can produce
macroscopic ensembles of highly aligned molecules under the
field-free condition. The mechanism of the field-free alignment is
well understood by the rephasing of the rotational wave packet
\cite{seideman,mette}. In detail, the target molecules are
impulsively excited by an intense ultrashort laser pulse into a
broad superposition of field-free rotational states. After the
alignment pulse is turned off, the excited wave packet will evolve
under the field-free condition and rephase to form a revival at
certain time delays. The revival structure in the nonadiabatic
alignment was first observed with the Coulomb explosion imaging by
Rosca-Pruna and Vrakking \cite{pruna}. Since then, the fundamental
behavior and dynamics of the alignment have been extensively
studied using the Coulomb explosion imaging and polarization
spectroscopy \cite{stapel,xu}. The investigation of nonadiabatic
alignment is motivated by several intriguing applications
including the use of alignment revivals to compress laser pulse
\cite{bartels}, the phase control of rotational wave packets
\cite{lee}, and its use to follow the time evolution of the
electron wave packets \cite{haessler}.

Recently it was demonstrated that the revival structure of
nonadiabatic alignment of molecules can also be observed via the
high-order harmonic generation (HHG) \cite{miyazaki,kanai}.
Therein, the evolution of the alignment parameter is imprinted in
the change of the harmonic yield with the delay between the
harmonic generating pulse and the alignment pulse. The modulation
due to the rotational dynamics, however, is influenced by the
symmetry of the molecules and the quantum interference in the HHG.
This may complicate the extraction of the information about the
rotational dynamics. In this paper, we theoretically investigate
the alignment-dependent spectral minimum in HHG from
nonadiabatically aligned CO$_2$, N$_2$, N$_2$O and CO molecules,
around their first half rotational revivals. It is found that the
alignment-dependent structural minimum presents a similar temporal
behavior to that of the alignment parameter $<\cos^2\theta>$ (i.e.
the expectation value of $\cos^2\theta$ where $\theta$ is the
angle between the molecular axis and the field direction) and this
phenomenon is not influenced by the change of the wavelength and
intensity of the probe pulse. To reveal the implication of the
phenomenon, we perform an analysis based on the two-center
interference model and find that the structural minimum position
depends linearly on the inverse of the alignment parameter. As the
molecular rotational dynamics is well characterized by the
evolution of $<\cos^2\theta>$, our results indicate the
possibility of tracing the molecular rotational dynamics by
measuring the spectral minimum.

\section{Theoretical model}

In our simulation, we use the strong-field approximation (SFA)
model for molecules \cite{lewen,qin1} to calculate the HHG
spectrum for a fixed alignment. Within the single active electron
(SAE) approximation, the time-dependent dipole velocity is given
by
\begin{eqnarray}
\mathbf{v}_{dip}(t;\theta) & = & i\int^t_{-\infty}
dt'\left[\frac{\pi}{\zeta+i(t-t')/2}\right]^{\frac{3}{2}}\exp[-iS_{st}(t',t)]\nonumber\\
 & &\times\mathbf{F}(t')\cdot\mathbf{d}_{ion}\left[\mathbf{p}_{st}(t',t)+\mathbf{A}(t');\theta\right]
 \nonumber\\
 &
 &\times\mathbf{v}_{rec}^{*}\left[\mathbf{p}_{st}(t',t)+\mathbf{A}(t);\theta\right]+c.c..
 \end{eqnarray}
In this equation, $\zeta$ is a positive constant. $t'$ and $t$
correspond to the ionization and recombination time of the
electron, respectively. $\theta$ is the alignment angle between
the molecular axis and the polarization of the probe pulse.
$\mathbf{F}(t)$ refers to the electric field of the probe pulse,
and $\mathbf{A}(t)$ is its associated vector potential.
$\mathbf{p}_{st}$ and $S_{st}$ are the stationary momentum and the
quasi-classical action, which are given by
\begin{equation}
\mathbf{p}_{st}(t',t)=-\frac{1}{t-t'}\int^t_{t'}\mathbf{A}(t'')dt''
\end{equation}
and
\begin{equation}
S_{st}(t',t)=\int^t_{t'}(\frac{[\mathbf{p}_{st}+\mathbf{A}(t'')]^2}{2}+I_p)dt''
\end{equation}
with $I_p$ being the ionization energy of the molecular state that
the electron is ionized from.
$\mathbf{d}_{ion}\left[p;\theta\right]$ and
$\mathbf{v}_{rec}\left[p;\theta\right]$ are the dipole matrix
elements describing the ionization and recombination of the
electron, respectively. Within the SFA model, the re-colliding
wave packet is approximately described by the plane wave. Hence
the dipole matrix elements are given by
\begin{equation}
\mathbf{d}_{ion}\left[p;\theta\right]=\langle\Psi_0(x,y,z;\theta)|\vec{r}|\Psi_p\rangle,
\end{equation}
\begin{equation}
\mathbf{v}_{rec}\left[p;\theta\right]=\langle\Psi_0(x,y,z;\theta)|-i\triangledown_r|\Psi_p\rangle,
\end{equation}
with $\Psi_p=\exp(i\vec{p}\cdot\vec{r})$ being the electronic
continuum state and $\Psi_0(x,y,z;\theta)$ being the ground state
of the target molecule. After the time-dependent dipole velocity
is obtained, the complex amplitude of the high-order harmonics
with a frequency $\omega_n$ is calculated by
\begin{equation}
\tilde{\mathbf{E}}(\omega_n;\theta)=\int e^{i\omega_n
t}\frac{d}{dt}\mathbf{v}_{dip}(t;\theta)dt.
\end{equation}

Coherently superposing the high-order harmonic emissions at
different alignment angles weighted by the angular distribution,
the spectrum at the delay $\tau$ with respect to the pump pulse is
obtained
\begin{equation}
S(\omega_n;\tau)=|\int
E(\omega_n;\theta)exp[iP(\omega_n;\theta)]\rho(\theta;\tau)d\theta|^2.
\end{equation}
Here, $E(\omega_n;\theta)=|\tilde{\mathbf{E}}(\omega_n;\theta)|$
and $P(\omega_n;\theta)=\arg(\tilde{\mathbf{E}}(\omega_n;\theta))$
are the amplitude and phase of the high-order harmonics generated
from the molecule aligned at $\theta$. $\rho(\theta;\tau)$ is the
angular distribution multiplied by $\sin\theta$ and is given by
\begin{equation}
\rho(\theta;\tau)=\sin\theta(1/Z)\sum_{J_i}Q(J_i)
\sum_{M_i=-J_i}^{J_i}\int|\Psi^{J_iM_i}(\theta,\phi;\tau)|^2
d\phi.
\end{equation}
Here $Q(J_i)=g_{J_i}\exp(-BJ_i(J_i+1)/(k_BT))$ is the Boltzmann
distribution function of the initial field-free state
$|J_i,M_i\rangle$ at temperature $T$,
$Z=\sum^{J_{max}}_{J=0}(2J+1)Q(J)$ is the partition function,
$k_B$ and $B$ are the Boltzmann constant and the rotational
constant of the molecule, respectively. $g_{J_i}$ is introduced to
include the different population of the initial J-states that
arises from the nuclear-spin statistics. For CO$_2$, only even-J
states are populated. For N$_2$, the population ratio between the
even- and odd-J states is 2:1, while for N$_2$O and CO, the
population ratio is 1:1. $\Psi^{J_iM_i}(\theta,\phi;\tau)$ is the
rotational wave packet excited from the initial state
$|J_i,M_i\rangle$, and is obtained by solving the time-dependent
Schr\"{o}dinger equation (TDSE) within the rigid-rotor
approximation \cite{ortigoso,seideman}
\begin{equation}
i\frac{\partial\Psi(\theta,\phi;\tau)}{\partial
\tau}=[B\mathbf{J}^2-\frac{E_p(\tau)^2}{2}(\alpha_\parallel
\cos^2\theta+\alpha_\perp \sin^2\theta)]\Psi(\theta,\phi;\tau).
\end{equation}
In this equation, $\alpha_\parallel$ and $\alpha_\perp$ are the
polarizabilities in parallel and perpendicular directions with
respect to the molecular axis, respectively. $E_p(\tau)$
represents the electric field of the alignment pulse. The degree
of alignment is characterized by the alignment parameter
$<\cos^2\theta>$, and is given by
\begin{equation}
<\cos^2\theta>(\tau) =(1/Z)\sum_{J_i}Q(J_i)
\sum_{M_i=-J_i}^{J_i}\langle\Psi^{J_iM_i}(\theta,\phi;\tau)|\cos^2\theta|\Psi^{J_iM_i}(\theta,\phi;\tau)\rangle.
\end{equation}

\section{Result and discussion}

In our simulation, linearly polarized alignment and probe pulses
are used, with their polarization directions parallel to each
other. In order to obtain a broadband spectrum to observe the
spectral minimum, the long-wavelength probe pulses are used. Under
these conditions, the influence of the multi-channel contribution
is small \cite{rupenyan} and can be neglected. Therefore only the
ionization channel involving the highest occupied molecular
orbital (HOMO) of the target is considered in our simulation.
Using Eqs. (1-6), the HHG spectrum for a fixed alignment is
calculated. The involved molecular orbital is calculated by using
the Gaussian 03 $ab\ initio$ code \cite{gauss} with the 6-311G
basis set for the CO$_2$, CO, N$_2$O molecules and with the 3-21G
basis set for the N$_2$ molecule. The rotational wave packet
excited by the alignment pulse is calculated by solving Eq. (9).
According to the rigid-rotor model, the rotational period of
molecules is determined by $T_r=1/(2Bc)$, with $B$ and c being the
rotational constant and the speed of the light, respectively. In
Table 1, the molecular parameters of the CO$_2$, N$_2$, N$_2$O,
and CO molecules used in the alignment calculation are listed.

\begin{center}
\small
\begin{tabular}{|p{2cm}|p{2cm}|p{2cm}|p{2cm}|p{2cm}|}
 \multicolumn {5}{c}
 {\bfseries TABLE 1: The molecular parameters of the CO$_2$,
N$_2$, N$_2$O, and CO molecules } \\
\hline
 \hline
  Molecule& B ($cm^{-1}$) & $\alpha_\parallel-\alpha_\perp$ (a.u.) & $\alpha_\perp$ (a.u.) & $T_{r}$ (ps) \\
 \hline
  CO$_2$ & 0.389 & 15.4 & 14.7 & 42.7 \\
  \hline
  N$_2$& 2.01 & 6.276 & 9.785 & 8.26   \\
  \hline
  N$_2$O & 0.41 & 18.85 & 13.16  & 40.5 \\
 \hline
  CO & 1.93 & 3.536 & 11.83 & 8.6 \\
  \hline
 \end{tabular}
 \end{center}

\begin{figure}[htb]
\centerline{\includegraphics[width=8.0cm]{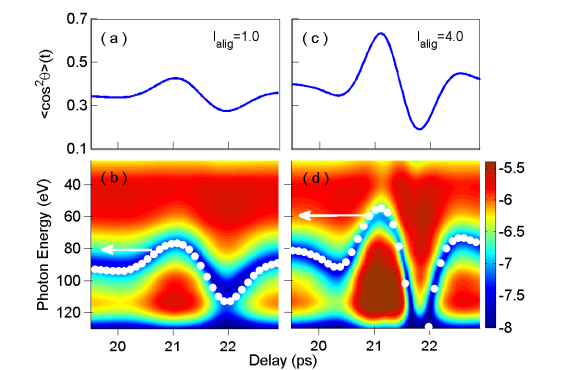}}
\setlength{\abovecaptionskip}{-0.1cm}
\setlength{\belowcaptionskip}{-0.12cm} \caption{(a-b) The
alignment parameter $<\cos^2\theta>$ (a) and the map of HHG
spectra (b) for the alignment pulse with an intensity of
$1.0\times10^{13}$ W/cm$^2$. (c-d) The alignment parameter
$<\cos^2\theta>$ (c) and the map of HHG spectra (d) for the
alignment pulse with an intensity of $4.0\times10^{13}$ W/cm$^2$.
The same probe pulse is used to obtain the two spectral maps. The
vertical axis representing the photon energy is upside-down. The
calculated photon energy of the spectral minimum is shown by the
white dots. The horizontal and vertical axes are the same as those
for the map of the HHG spectra.}
\end{figure}
We first simulate the alignment-dependent HHG spectrum generated
from CO$_2$ molecule. In Figs. 1(a) and 1(b), the evolution of the
alignment parameter $<\cos^2\theta>$ around the first half revival
(panel a) and the corresponding map of the HHG spectrum as a
function of the time delay and the photon energy (panel b) are
presented. For the map of the HHG spectrum (panel b), the
horizontal axis represents the time delay with respect to the
alignment pulse, and the vertical axis represents the photon
energy of the high-order harmonics. Here, an 800 nm, 100 fs
alignment pulse with an intensity of $1.0\times10^{13}$ W/cm$^2$
is used. The rotational temperature is set to be 40 K. The
wavelength, pulse duration, and intensity of the probe pulse used
to generate high-order harmonics are 1450 nm, 18 fs, and
$1.7\times10^{14}$ W/cm$^2$, respectively. In experiment, this IR
pulse can be produced by optical parametric amplification (OPA)
technology based the wavelength down-conversion of Ti:sapphire
laser source \cite{cerullo,zhang}. As shown in Fig. 1(b), obvious
spectral minimum is observed in the HHG spectrum at every time
delay. Moreover, the alignment-dependent spectral minimum has a
similar shape to that of the evolution of $<\cos^2\theta>$. In
detail, when the value of $<\cos^2\theta>$ increases (or
decreases), the spectral minimum position correspondingly shifts
to lower (or higher) photon energy \cite{rupenyan,jin}. The lowest
and highest photon energy of the spectral minima appear at the
delays when $<\cos^2\theta>$ reaches its maximum and minimum,
respectively. For a further investigation, we also study the cases
where the molecular alignment is achieved by using the pulses with
different intensities. The results for the $4.0\times10^{13}$
W/cm$^2$ alignment pulse are presented in the second column of
Fig. 1. As shown in Figs. 1(c) and 1(d), the evolution of the
spectral minimum also presents a similar shape to that of
$<\cos^2\theta>$. In this case, the maximum of $<\cos^2\theta>$ is
around 0.6. The corresponding spectral minimum is observed around
60.5 eV, which differs from the measurements in Ref.
\cite{rupenyan} by about 13 eV (close to the ionization energy of
CO$_2$ molecule). The disagreement arises from the plane-wave
approximation of the electronic continuum state in the SFA
\cite{haessler}. This is improved in the quantitative
re-scattering (QRS) theory by employing a more precise transition
dipole with scattering wave instead of the plane wave for the
recombination step in a HHG process \cite{le}. As shown in Figs.
1(a) and 1(c), when varying the intensity of the alignment pulse,
not only the amplitude but also the shape of $<\cos^2\theta>$ are
changed. The corresponding changes in the evolution of the
spectral minimum are also observed, as shown in Figs. 1(b) and
1(d). Hence, the shape of the alignment-dependent spectral minimum
varies with that of $<\cos^2\theta>$.

\begin{figure}[b]
\centering\includegraphics[width=7.0cm]{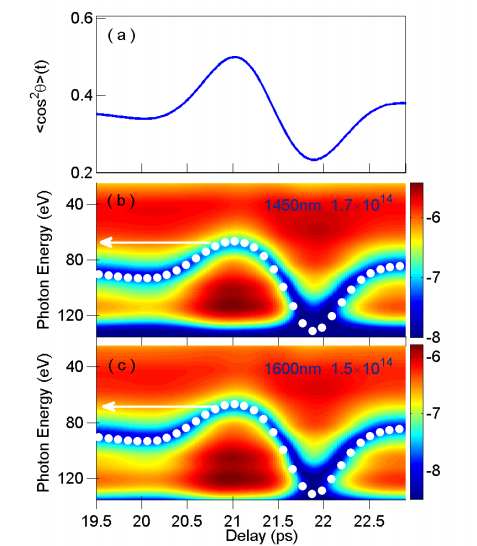}
\setlength{\abovecaptionskip}{-0.1cm}
\setlength{\belowcaptionskip}{-0.3cm}\caption{(a) The alignment
parameter $<\cos^2\theta>$ for the alignment pulse with an
intensity of $2.0\times10^{13}$ W/cm$^2$. (b-c) The maps of the
HHG spectra produced by two different probe pulses. The calculated
photon energy of the spectral minimum is shown by the white dots.
The horizontal and vertical axes are the same as those for the
maps of the HHG spectra.}
\end{figure}

In the following, we investigate the role of the probe pulse in
the phenomenon observed in Fig. 1. The maps of the harmonic
spectrum produced by two different probe pulses are displayed in
Fig. 2. The wavelength and intensity of the two probe pulses are
different, which are indicated in the corresponding maps of the
high-order harmonic spectrum. The same alignment pulse with an
intensity of $2.0\times10^{13}$ W/cm$^2$ is used for the two maps.
The alignment parameter $<\cos^2\theta>$ is presented in Fig.
2(a). Other parameters of the alignment and probe pulses are the
same as those used for Fig. 1. As shown in Fig. 2, the shapes of
the spectral minimum for the two different probe pulses are the
same, and simultaneously similar to that of $<\cos^2\theta>$.
Therefore, the imprint of $<\cos^2\theta>$ on the
alignment-dependent spectral minimum is not influenced by the
changing of the wavelength and intensity of the probe pulse.

\begin{figure}[b]
\centering\includegraphics[width=11.0cm]{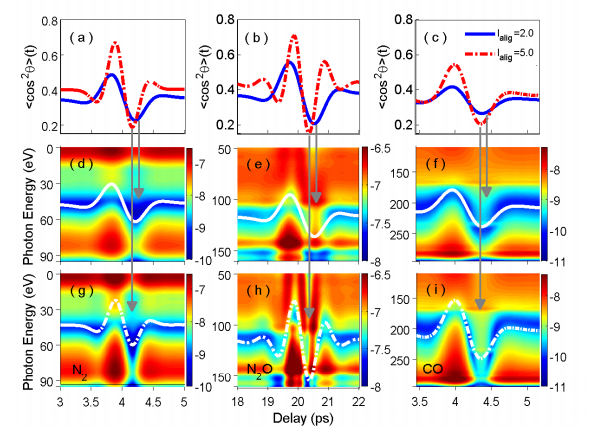}
\setlength{\abovecaptionskip}{-0.0cm}
\setlength{\belowcaptionskip}{-0.0cm} \caption{(a-c) The alignment
parameter $<\cos^2\theta>$ obtained by the $2.0\times10^{13}$
W/cm$^2$ (blue solid line) and $5.0\times10^{13}$ W/cm$^2$ (red
dash-dotted line) alignment pulse. (d-f) The maps of the
high-order harmonic spectra for the $2.0\times10^{13}$ W/cm$^2$
alignment pulse. (g-i) The maps of the high-order harmonic spectra
for the $5.0\times10^{13}$ W/cm$^2$ alignment pulse. The first,
second, and third columns present the results for N$_2$, N$_2$O
and CO molecules, respectively. For clarity, the curves of
$<\cos^2\theta>$ are depicted by the white solid line for the
$2.0\times10^{13}$ W/cm$^2$ alignment pulse and by the white
dash-dotted line for the $5.0\times10^{13}$ W/cm$^2$ alignment
pulse on the corresponding maps of the high-order harmonic
spectra.}
\end{figure}

To get an insight into the relation between the spectral minimum
and the alignment parameter $<\cos^2\theta>$, we perform an
analysis based on the two-center interference model
\cite{kanai,lein,zhu1}. According to this model, the condition for
the interference minimum is $R\cos\theta=n\lambda$ for molecules
with an anti-bounding symmetric HOMO. Here $\lambda$ is the de
Broglie wavelength of the returning electron, $R$ is the
internuclear distance of the target molecule, and n is a positive
integer. According to the relationship $\omega=\frac{k^2}{2}+I_p$
with the electron momentum $k=\frac{2\pi}{\lambda}$ (atomic units
are used), one can obtain
$\omega=\frac{2\pi^2}{R^2\cos^2\theta}+I_p$, where $\omega$ and
I$_p$ are the photon energy of the interference minimum and the
ionization energy of the target molecule, respectively. The above
formula is derived based on the responses of the molecules aligned
at a fixed angle. After taking into account the partial alignment,
this relation between the minimum position and $\cos^2\theta$ can
still remain, i.e.
$\tilde{\omega}=\frac{2\pi^2}{R^2}\frac{1}{<\cos^2\theta>}+I_p$.
Here, $\tilde{\omega}$ represents the minimum position in the
high-order harmonic spectrum generated from partially aligned
molecules, $<\cos^2\theta>$ is the alignment parameter. As for the
molecule with a bounding symmetric HOMO, the similar formula is
given by
$\tilde{\omega}=\frac{\pi^2}{2R^2}\frac{1}{<\cos^2\theta>}+I_p$.
We apply the above formula to the CO$_2$ molecule, whose HOMO
possesses anti-bounding symmetry, and calculate the photon energy
of the spectral minimum with the value of $<\cos^2\theta>$. The
results are presented by the white dots in Fig. 1 and Fig. 2. The
horizontal and vertical axes are the same as those for the map of
the high-order harmonic spectrum. As shown by the white dots in
Fig. 1 and Fig. 2, the calculated photon energy of the spectral
minimum agrees well with that observed in the high-order harmonic
spectrum. Hence the alignment-dependent spectral minimum position
is well reproduced by the formula with $<\cos^2\theta>$ and
depends linearly on the inverse of the alignment parameter
$\frac{1}{<\cos^2\theta>}$.

To explore the universality of the phenomenon observed in the
CO$_2$ molecule, we investigate the alignment-dependent spectral
minimum of N$_2$, N$_2$O and CO molecules that possess different
orbital symmetry. For each molecule, two alignment pulses with
different intensities, i.e. $2.0\times10^{13}$ W/cm$^2$ and
$5.0\times10^{13}$ W/cm$^2$, are considered. Other parameters of
the alignment pulses and the rotational temperature are the same
as those used in Fig. 1 and Fig. 2. In the case of N$_2$ molecule,
a 1600 nm, 20 fs probe pulse with an intensity of
$0.9\times10^{14}$ W/cm$^2$ is used. For N$_2$O molecule, the
wavelength, pulse duration, and intensity of the probe pulse are
2000 nm, 15 fs, and $1.2\times10^{14}$ W/cm$^2$, respectively. In
the CO molecule case, because the spectral minimum occurs at
higher photon energy \cite{zhu1}, a probe pulse with a longer
wavelength is used. The wavelength, pulse duration, and intensity
are 2500 nm, 25 fs, and $1.5\times10^{14}$ W/cm$^2$, respectively.
In our simulation, few-cycle long-wavelength probe pulses with low
intensities are used for all the molecules. Hence the influence of
the multi-channel contribution in HHG is small and can be
neglected \cite{vozzi,rupenyan}.

In Figure. 3, the alignment parameter $<\cos^2\theta>$ and the
corresponding maps of the high-order harmonic spectrum for N$_2$,
N$_2$O and CO molecules are presented in the first, second, and
third columns, respectively. From Fig. 3, one can see that,
similar to the case of CO$_2$ molecule, the evolution of the
alignment parameter $<\cos^2\theta>$ is also imprinted in the
corresponding spectral minimum for the three molecules. The
temporal behavior of the spectral minimum varies with that of
$<\cos^2\theta>$. For clarity, the curves of $<\cos^2\theta>$ are
depicted by the white lines in the corresponding maps of the
high-order harmonic spectrum. As shown in the second and third
rows of Fig. 3, the white lines representing $<\cos^2\theta>$ and
the minimum position have similar temporal behavior. One can also
see that the white lines do not always match the position of the
spectral minimum, which is very obvious in the case of the
stronger degrees of alignment shown in Figs. 3(g-i). These
deviations are due to the fact that the spectral minimum position
depends linearly on $\frac{1}{<\cos^2\theta>}$ rather than
$<\cos^2\theta>$, which will be demonstrated in Fig. 4. Compared
with the cases of the N$_2$ and CO molecules shown in the first
and third columns of Fig. 3, the structures of the high-order
harmonic spectral maps of N$_2$O molecule shown in the second
column of Fig. 3 are more complicated and the spectral minimum is
difficult to identify. As demonstrated in the recent work
\cite{weber}, this is due to the intricate evolution of the
rotational wave packet shown in Fig. 3(b) and the complicated
structure of the recombination dipole.

\begin{figure}[htb]
\centering\includegraphics[width=10cm]{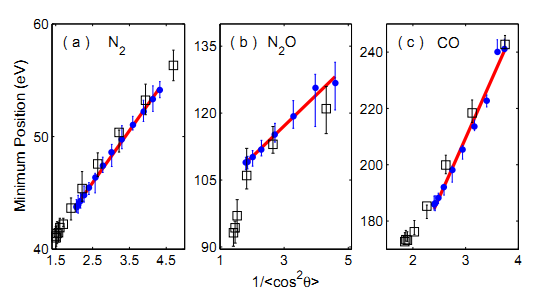}
\setlength{\abovecaptionskip}{-0.0cm}
\setlength{\belowcaptionskip}{-0cm} \caption{The extracted
spectral minimum positions from the second (the blue dots) and
third (the black squares) rows of Fig. 3 and the corresponding
least-square fits (red solid line) for the N$_2$ (panel a), N$_2$O
(panel b) and CO (panel c) molecules. The horizontal axis
represents the inverse of the alignment parameter.}
\end{figure}
We then apply the formulae derived from two-center interference
model to the N$_2$, N$_2$O and CO molecules. Whereas the
calculated spectral minimum positions do not agree well with those
observed in the high-order harmonic spectral maps (the results are
not shown here). This is because for N$_2$ molecule, its HOMO
exhibits a strong mixing between $s$ and $p$ states, while for
N$_2$O and CO molecules with nonsymmetric orbitals, the components
of the HOMO become more complicated. As a result, the spectral
minimum position before taking into account the partial alignment
does not exactly agree with that predicted by the two-center
interference model. Hence, the formulae can not quantitatively
describe the relation between the minimum position and the
alignment parameter. Even so, the linear relation between
$\tilde{\omega}$ and $\frac{1}{<\cos^2\theta>}$ still remains with
the two linear coefficients to be determined, considering the
similarity of the temporal behavior between the spectral minimum
position and $<\cos^2\theta>$ also observed in the cases of the
N$_2$, N$_2$O and CO molecules. To determine the two linear
coefficients, we fit the spectral minimum positions observed in
the maps of the high-order harmonic spectrum shown in the second
row of Fig. 3, through the linear least-squares fitting method.
The results are presented in Fig. 4 for N$_2$ (panel a), N$_2$O
(panel b) and CO (panel c). The minimum positions extracted from
the second and third rows of Fig. 3 are presented by the blue dots
and the black squares, respectively. The error bars of the data
points, which describes how well the spectral minimum can be
defined from Fig. 3, are also depicted. As shown in Fig. 4, the
extracted spectral minima from the second row of Fig. 3 are well
fitted by a linear function of $\frac{1}{<\cos^2\theta>}$ for the
three molecules. Furthermore, for the N$_2$ and CO molecules, the
minimum positions extracted from the third row of Fig. 3 follow
the same linear function of $\frac{1}{<\cos^2\theta>}$ as those
extracted from the second row of Fig. 3. As for the N$_2$O
molecule, however, the spectral minimum position for the stronger
degree of alignment does not follow the same function of
$\frac{1}{<\cos^2\theta>}$ as those extracted from Fig. 3(e). The
relationship between the minimum position and
$\frac{1}{<\cos^2\theta>}$ is even not linear, as shown by the
black squares in Fig. 4(b). This is because the HOMO of the N$_2$O
molecule is contributed by three atoms, which may complicate the
relationship between the spectral minimum position and
$\frac{1}{<\cos^2\theta>}$. Therefore, when the molecular orbital
is contributed by two atomic centers, the structural minimum
position $\tilde{\omega}$ depends linearly on the inverse of the
alignment parameter $\frac{1}{<\cos^2\theta>}$. The linear
coefficients are independent on the alignment and probe pulses, as
shown in Figs. 1, 2 and 4. Hence, after determining the linear
coefficients for the target molecules, the formula can be used to
trace the evolution of $<\cos^2\theta>$ with the measured
structural minima in the experiments using various alignment and
probe pulses. As the molecular rotational wave-packet dynamics is
well characterized by the alignment parameter $<\cos^2\theta>$,
one can trace the rotational wave-packet dynamics by measuring the
alignment-dependent spectral minima. Compared with the Coulomb
explosion imaging, our method can realize all-optical measurement
of the angular distribution without destroying the molecules and
can directly trace the value of $<\cos^2\theta>$ through a simple
formula of the spectral minimum. The precision of tracing the
rotational dynamics with our method depends on the identification
of the minimum position. When the spectral minimum is difficult to
identify, correspondingly, it will be difficult to precisely trace
the rotational dynamics.

\section{Conclusion}
In summary, we investigate the dependence of the spectral minimum
position on the alignment parameter $<\cos^2\theta>$ around the
first half rotational revival for the CO$_2$, N$_2$, N$_2$O and CO
molecules. It is found that the alignment-dependent spectral
minimum presents a similar temporal behavior to that of
$<\cos^2\theta>$. Our analysis shows that the minimum position
depends linearly on the inverse of $<\cos^2\theta>$. For the
molecular orbitals contributed by two atomic centers, the two
linear coefficients are independent on the alignment and probe
pulses. As the molecular rotational dynamics is well characterized
by the evolution of $<\cos^2\theta>$, it is possible to probe the
rotational wave-packet dynamics by measuring the
alignment-dependent spectral minima.

\section*{Acknowledgment}
This work was supported by the NNSF of China under Grants No.
11234004 and 61275126, the 973 Program of China under Grant No.
2011CB808103 and the Doctoral fund of Ministry of Education of
China under Grant No. 20100142110047.

\end{document}